\documentclass[aps,superscriptaddress, twocolumn,prl, showpacs,preprintnumbers,nofootinbib, amsmath,amssymb,
aps, superscriptaddress]{revtex4-1}

\usepackage{graphicx}
\usepackage{amsmath,amssymb}
\usepackage{amsfonts}
\usepackage{xspace} 
\usepackage[usenames]{color}
\usepackage{dcolumn}
\usepackage{bm}
\usepackage{mathrsfs}
\usepackage{multirow}
\usepackage[colorlinks=true]{hyperref}
\usepackage[all]{hypcap} 
\usepackage[utf8]{inputenc} 
\usepackage{lipsum}
\usepackage{etoolbox}
\usepackage{tikz}
\usepackage{url}
\usepackage{subfigure}
\usepackage{adjustbox}
\usepackage{caption}
\usepackage{slashed}
\usepackage{multirow}
\usepackage{array}
\usepackage{booktabs}
\usepackage{siunitx}
\usepackage{comment}
\usepackage{lineno}
\usepackage{orcidlink}

\def\be{\begin{equation}}
\def\ee{\end{equation}}
\def\bea{\begin{eqnarray}}
\def\eea{\end{eqnarray}}

\newcommand{\beq}{\begin{eqnarray}}
\newcommand{\eeq}{\end{eqnarray}}
\usepackage{calrsfs}
\DeclareMathAlphabet{\pazocal}{OMS}{zplm}{m}{n}

\newcommand*\Bell{\ensuremath{\boldsymbol\ell}}

\begin{document}

\title{Electromagnetic duality anomaly in accelerating  waveguides}

\author{Adri\'an del R\'{\i}o\,\orcidlink{0000-0002-9978-2211}}\email{adrdelri@math.uc3m.es}
 \affiliation{Universidad Carlos III de Madrid, Departamento de Matem\'aticas.\\ Avenida de la Universidad 30 (edificio Sabatini), 28911 Legan\'es (Madrid), Spain.}
 
\begin{abstract}

The classical symmetry of the source-free Maxwell equations under electric-magnetic duality rotations leads to a conserved Noether charge,  corresponding to  the circular polarization of light. We show that, in quantum field theory, the vacuum expectation value of this charge is no longer time-independent inside a long, cylindrical waveguide undergoing both linear and rotational acceleration from rest. Specifically, photon pairs are spontaneously excited from the quantum vacuum by the accelerated background, while the helical motion induces an imbalance in the number of right- and left-handed modes produced. This photon helicity non-conservation reflects a genuine quantum effect that breaks the classical duality symmetry. The model considered  paves the way for experimental studies.

\end{abstract}
\maketitle

\noindent\textit{\textbf{Introduction}.}  Maxwell's equations in vacuum exhibit an electric-magnetic duality invariance. This is a Noether symmetry of the source-free classical  theory in any curved spacetime \cite{PhysRevD.13.1592}, and its associated conserved charge quantifies the difference between right- and left-handed  light intensities \cite{10.1119/1.1971089}, corresponding to the Stokes V parameter in optics. As a result, the circular polarization state of classical electromagnetic waves is a constant of motion, even in the presence of strong gravitational fields.

Remarkably, quantum vacuum fluctuations of the electromagnetic field, when enhanced by gravitational effects, can break this symmetry  \cite{PhysRevLett.118.111301, PhysRevD.98.125001}. Specifically, the vacuum expectation value of the Noether charge in the quantum theory is no longer conserved, and its time evolution is dictated by the spacetime geometry. When this occurs, the symmetry is deemed anomalous. This result can be understood as the spin-1 analogue of the well-known chiral anomaly for fermions in QED \cite{PhysRev.177.2426, 1969NCimA..60...47B}. Interestingly, this anomaly arises if and only if the gravitational field carries a flux of circularly polarized gravitational waves \cite{PhysRevLett.124.211301, PhysRevD.104.065012}, such as those produced in binary black hole mergers with a preferred helicity direction \cite{PhysRevD.108.044052, leong2025gravitationalwavesignaturesmirrorasymmetry}. As a consequence of this quantum effect, photon pairs are expected to be spontaneously excited from the quantum vacuum by an external gravitational field, violating helicity conservation and generating an imbalance in the number of right- and left-handed photons  \cite{doi:10.1142/S0218271817420019, sym10120763}.

The electromagnetic duality inspired Dirac to postulate the existence of magnetic monopoles as a way to explain charge quantization \cite{Dirac1931kp}, and its discrete counterpart plays an important role in gauge and string theories by relating the strong and weak coupling regimes \cite{SEIBERG199419}. Despite the profound conceptual implications of this anomaly in quantum field theory, its observable consequences in realistic astrophysical scenarios are expected to be negligible. Indeed,  results from \cite{PhysRevLett.124.211301} suggest that the net photon helicity generated from the quantum vacuum during solar-mass binary black hole  mergers, is of the same order of magnitude as  Hawking radiation \cite{cmp/1103899181}. This motivates the search for manifestations of the electric-magnetic duality anomaly in analogue-gravity systems---a growing field that enables laboratory simulations of quantum field effects in curved spacetimes, typically  using condensed matter systems or photonic platforms \cite{Barcel__2005}.

According to the Equivalence Principle, a gravitational field is  indistinguishable from accelerated motion as measured by local observers. Therefore, a similar anomaly is expected to occur in accelerated cylindrical waveguides with a preferred helicity direction. 
 In this Letter, I present a toy model that realizes this quantum effect, thereby opening a path toward laboratory studies of the duality anomaly in more realistic settings. In contrast to  \cite{PhysRevLett.118.111301, PhysRevD.98.125001}, the approach here provides a direct derivation of the vacuum expectation value (VEV), avoids formal computational methods, and explicitly demonstrates the connection between the anomaly and asymmetric photon pair production. Throughout this work, we adopt geometric units $G = c = 1$, while keeping Planck's constant $\hbar\neq 1$ explicit to highlight quantum effects.

\noindent\textit{\textbf{The setup: accelerating waveguide and  reference frames}.} 
We consider an infinitely long, hollow cylindrical waveguide  of radius $R$ in Minkowski spacetime $(\mathbb R^4,\eta_{ab})$, where $\eta_{ab}$ denotes the flat  metric. Initially static, the cylinder accelerates both tangentially and longitudinally until it reaches a constant angular velocity $\Omega_0$ and constant linear velocity $v_0$ along its  symmetry axis. To solve Maxwell's equations,  it is convenient to work in a fiducial reference frame $\tilde O$ associated with observers co-rotating and co-propagating with the waveguide, so that   boundary conditions remain ``time-independent''. In cylindrical coordinates $\{\tilde t,\tilde \rho,\tilde\phi,\tilde z\}$ adapted to this frame,  the flat Minkowski metric reads $ds^2= \tilde \eta_{ab}( \tilde x) d \tilde x^a d \tilde x^b=-d\tilde t^2+d\tilde \rho^2+\tilde\rho^2d\tilde\phi^2+d\tilde z^2$, and each spacetime point carries a natural  orthonormal basis of  vector fields $\{\tilde t_a, \tilde\rho_a, \tilde \phi_a, \tilde z_a\}=\{-\nabla_a \tilde t, \nabla_a \tilde \rho, \tilde \rho\nabla_a \tilde \phi, \nabla_a \tilde z\}$, where $\tilde t^a$ is  the 4-velocity of the co-moving observers.  In this frame, points $p$ of  the waveguide remain at rest, with their spatial coordinates $\{\tilde \rho(p),\tilde\phi(p),\tilde z(p)\}$ constant in time $\tilde t$.

Now, we are interested  in describing the quantum field from the viewpoint of a reference frame $O$ decoupled from the waveguide, and attached to inertial observers that see the waveguide accelerate and, at late times, rotate and move uniformly. Let $\{t,\rho,\phi,z\}$ denote cylindrical coordinates which remain constant along the worldlines of  observers at rest in this reference frame. At early times, both frames coincide. However, at late times the two coordinate systems are related by $\rho=\tilde \rho$, $\phi=\tilde\phi-\Omega_0 \tilde t$ (time-dependent rotation),  $z=\gamma(\tilde z+v_0 \tilde t)$, $t=\gamma(\tilde t+v_0 \tilde z)$ (Lorentz transformation with boost parameter $\gamma^{-2}=1-v_0^2$ ), and the fiducial, co-moving tetrad becomes a moving frame when viewed by inertial observers in $O$:  
\bea
t_a|_{(t,\rho,\phi,z)}&=&\cosh \theta \, \, \tilde t_a|_{(\tilde t,\tilde\rho,\tilde\phi,\tilde z)} - \sinh \theta \,\, \tilde z_a|_{(\tilde t,\tilde\rho,\tilde \phi,\tilde z)} \nonumber\, ,\\
\rho_a|_{(t,\rho,\phi,z)}&=&\cos \Omega_0 \tilde t\,  \tilde \rho_a|_{(\tilde t,\tilde\rho,\tilde\phi,\tilde z)} + \sin \Omega_0 \tilde t \, \tilde \phi_a|_{(\tilde t,\tilde\rho,\tilde \phi,\tilde z)} \nonumber\, ,\\
\phi_a|_{(t,\rho,\phi,z)}&=&\cos \Omega_0 \tilde t\, \tilde  \phi_a|_{(\tilde t,\tilde\rho,\tilde\phi,\tilde z)} - \sin \Omega_0 \tilde t \, \tilde \rho_a|_{(\tilde t,\tilde\rho,\tilde \phi,\tilde z)} \nonumber\, ,\\
z_a|_{(t,\rho,\phi,z)}&=&\cosh \theta\, \, \tilde z_a|_{(\tilde t,\tilde\rho,\tilde\phi,\tilde z)} - \sinh \theta \, \, \tilde t_a|_{(\tilde t,\tilde\rho,\tilde \phi,\tilde z)}  \nonumber\, ,
\eea
 where $\cosh\theta\equiv\gamma$, $\sinh\theta\equiv v_0\gamma$.

\noindent\textit{\textbf{The classical duality symmetry and its Noether charge inside the waveguide}.}   Let $\nabla_a$ denote the   covariant derivative associated with the flat  metric $\eta_{ab}$ (Levi-Civita connection). The dynamics of the electromagnetic field $F_{ab}$ is governed by the source-free Maxwell equations, $\nabla_a F^{ab}=0$,  $\nabla_a {^*F^{ab}}=0$ ---where $*$ denotes the Hodge dual operator--- along with appropriate boundary conditions  and suitable initial data on a spacelike Cauchy hypersurface $\Sigma$. As one can easily check, the field equations are invariant under $F_{ab}\to \cos \theta F_{ab}+\sin \theta \, {^*F}_{ab}$, ${^*F}_{ab}\to \cos \theta {^*F}_{ab}-\sin \theta \, {F}_{ab}$, $\forall\theta\in\mathbb R$, known as an electric-magnetic duality rotation.  The   specific form of the boundary conditions depends on the physical nature of the waveguide,  effectively modeling the interaction of the electromagnetic field with the microscopic constituents of this background. For instance, in a perfectly conducting cylinder, the electric field is normal to the surface while the magnetic one is tangential.   In our problem, we will study instead a toy model where the boundary conditions also remain invariant under an electric-magnetic duality rotation. Specifically, in the fiducial co-moving frame we  demand $\tilde z^a \tilde E_a\equiv \tilde t^a \tilde z^b  \tilde F_{ab}=0$, $\tilde z^a  \tilde B_a\equiv \tilde t^a \tilde z^b {^{\tilde *} \tilde F}_{ab}=0$ on the boundary $\partial C$. While not strictly necessary (see the discussion section), this choice  makes the calculation of the quantum anomaly more transparent.

In terms of a (magnetic) potential $A_a$, defined by $F_{ab}=\nabla_a A_b-\nabla_bA_a$, the space of solutions to Maxwell equations, Sol($\mathbb S$) $=\{A_a\in \Lambda^1(\mathbb R^4)\, / \, \nabla_a F^{ab}=0,\,   \tilde t^a \tilde z^b \tilde F_{ab}  |_{\partial C}=\tilde t^a \tilde z^b {^*\tilde F}_{ab}|_{\partial C}=0\}$, can be endowed with a canonical symplectic structure 
\bea
\Omega_0(A_1,A_2)=\frac{1}{2}\int_\Sigma d\Sigma  \,  n_b\left[A_{a,1}F^{ba}_2 - A_{a,2} F_1^{ba} \right]  \label{symp0}\, ,
\eea
where $n_a=-\nabla_a t$ is the unit normal to $\Sigma$.
This geometric structure on Sol($\mathbb S$) provides a powerful framework for studying  the dynamics and symmetries of the physical system, and for identifying conserved Noether quantities \cite{10.1063/1.528801, PhysRevD.103.025011}.  First, for any two solutions to our problem, $A_1, A_2\in $ Sol($\mathbb S$), it can be shown that the integral above is independent of  time $t$ given the specified boundary conditions, thus its value is independent of the choice of $\Sigma$.  
Second,  consider a transformation rule, i.e. a map $S$: Sol($\mathbb S)\to  \Lambda^1(\mathbb R^4)$, which can be continuosly deformed around the identity, $S[A]=A+\delta_S A +...$. If $S$ is a symmetry of both the field equations {\it and} boundary conditions, this is, if $S[A]\in $ Sol($\mathbb S$) and therefore $\delta_S A\in $ Sol($\mathbb S$), then the value $Q(A):=\Omega_0(A,\delta_S A)$ is a constant of motion by virtue of the time-independence of $\Omega_0$. This is precisely the Noether charge that generates the symmetry on Sol($\mathbb S$). 

An electric-magnetic duality rotation can be  generated by the infinitesimal transformation $\delta_S A= Z$, where $Z$ is the ``dual'' (electric) potential that solves the Gauss constraint $D_a E^a=0$ in the source-free theory  \cite{PhysRevLett.118.111301, PhysRevD.98.125001} (similarly, $A_a$ solves $D_a B^a=0$). The resulting Noether charge reads 
\bea
Q_0(A)=\Omega_0(A,Z)=\frac{1}{2}\int_\Sigma d\Sigma\,  n_b\left[ A_a {^*F}^{ba}  -  Z_a F^{ba} \right].
\eea
Because the boundary conditions were chosen to remain invariant under electric-magnetic rotations, both $A_a$ and $Z_a$ belong to Sol($\mathbb S$),  ensuring that this expression is constant in time inside the accelerating waveguide. This can be explicitly checked by directly computing  $dQ_0/dt=0$. We note that  $Q_0(A)$ need not be conserved with other boundary conditions, even if the field equations themselves are duality invariant.

\noindent\textit{\textbf{Covariant phase space in the classical theory}.}
To understand the physical significance of $Q(A)$, and to eventually  do the quantization, we need to specify the elements of the covariant phase space Sol($\mathbb S$). This requires fixing a gauge choice, as well as a suitable ``representation'' of Maxwell solutions.  A particularly useful choice of variables is $A_a^R=\frac{1}{\sqrt{2}}[A_a+iZ_a]$, $A_a^L=\frac{1}{\sqrt{2}}[A_a-iZ_a]$, corresponding to self-dual (+)  and anti self-dual (-) fields $F_{ab}^{\pm}=\frac{1}{\sqrt{2}} [F_{ab}\pm i {^*F_{ab}}]$, respectively, which satisfy $i^* F_{ab}^{\pm}=\pm F_{ab}^{\pm}$. These complex variables ``diagonalize'' the field equations and the boundary conditions, which decouple and can then  be solved independently more easily. On the other hand, we will fix the Lorentz gauge $\nabla_a A_R^a=0$, as the  identification of photon helicity  becomes more transparent than in other choices.

In the fiducial comoving frame, it can be shown  that any solution of the source-free Maxwell equations inside the waveguide, with the specified boundary conditions, can be expressed as a linear combination
\bea\label{generalsolution}
A^R_a&=&\sum_{n=1}^{\infty} \sum_{m=-\infty}^{\infty}\int_{-\infty}^{\infty} dk\left[ z_{ k m n}^{R, \rm in}A_{a,1kmn}^{R, \rm in}+ \overline{ z_{ k m n}^{L, \rm in}} \, \overline{ A_{a,1kmn}^{L, \rm in}} \right]\, ,  
\eea
where $\{z_{ k m n}^{R, \rm in}, z_{ k m n}^{L, \rm in}\}$ are some complex-valued constants, and the circular polarization vector basis are (detailed computations will be published elsewhere)
\bea
A_{a,hkmn}^{R, \rm in}=\frac{i R^2 }{j_{mn}^2}e^{-i(h\omega_{kmn}\tilde t+k\tilde z+m\tilde \phi)}\hspace{4cm} \nonumber\\
\times \left[  \frac{j_{mn}}{R} J_{m+1}\left(  \frac{j_{mn}}{R}\tilde \rho\right) {\bf \overline{ \tilde m_a}}-i (h\omega_{kmn}+k) \tilde{\bf \Bell_a} J_{m}\left(  \frac{j_{mn}}{R}\tilde \rho\right) \right] \, , \nonumber\\
A_{a,hkmn}^{L, \rm in} =\frac{i R^2 }{j_{mn}^2}e^{-i(h\omega_{kmn}\tilde t+k\tilde z+m\tilde \phi)} \hspace{4cm}\nonumber\\
\times \left[   \frac{j_{mn}}{R} J_{m-1}\left(  \frac{j_{mn}}{R} \tilde\rho\right)  {\bf \tilde m_a}+i (h\omega_{kmn}+k)\tilde{\bf \Bell_a} J_{m}\left(  \frac{j_{mn}}{R} \tilde\rho\right) \right] \nonumber\, ,
\eea
where  $k\in \mathbb R$ and  $m\in \mathbb Z$  represent the linear and angular momenta, respectively, of the electromagnetic waves along the $z$ direction; $\omega_{kmn}=\sqrt{k^2+\frac{j^2_{mn}}{R^2}}$ are the allowed oscillation frequencies inside the waveguide; $h\in \{+1,-1\}$ is related to the wave's helicity; and $\{j_{mn}\}_{n=1,2,\dots}$ are the zeroes of the Bessel $J_m(x)$ function.  The modes $A_{a,hkmn}^{R, \rm in}$ ($A_{a,hkmn}^{L, \rm in}$) are  right(left)-handed, meaning that the their field strength is self-dual (anti self-dual): $\frac{1}{2}\epsilon_{ab}^{\quad cd}\nabla_{[c}A^{R/L, \rm in}_{d],h k m n} = \pm i \nabla_{[a}A^{R/L, \rm in}_{b],h k m n}$. Both chiral field modes are related by $\overline{A_{a,-h-k-mn}^{L, \rm in}}=(-1)^m A_{a,hkmn}^{R, \rm in}$.

The circular polarization vector basis is expressed in terms of the cylindrical Newman-Penrose null tetrad ${\bf \{ \tilde n_a,\tilde\Bell_a, \tilde m_a, \overline{\tilde m_a}\}}=\frac{1}{\sqrt{2}}\{\tilde t_a +\tilde z_a,  \tilde t_a -  \tilde z_a,\tilde\rho_a- i  \tilde\phi_a, \tilde\rho_a+ i  \tilde\phi_a\}$ \cite{Torres2003}.  The vector ${\bf \tilde \Bell_a}$ denotes the null ``outgoing'' direction of propagation along the $z$ axis, while ${\bf { \tilde m_a}}$ indicates the direction of rotation in the plane orthogonal to it.  For positive-frequency modes, $A_{a,1kmn}^{R,\rm in}$ rotates right-handedly along its propagation direction (thus, we say it has positive helicity), while for negative-frequency modes it rotates left-handedly  (thus, negative helicity). Similarly, $A_{a,1kmn}^{L, \rm in}$ rotates left-handedly along its propagation direction for positive-frequency modes (thus, negative helicity), while for negative-frequency modes it rotates right-handedly along its propagation direction (thus, negative helicity). This is the standard relation between duality and helicity \cite{10.1063/1.527187}.

The  covariant phase space in the classical theory will consist of the (real) vector space $\Gamma_R$  spanned by these (complex-valued) basis solutions. Each element in this vector space is labelled by $\{z_{ k m n}^{R, \rm in}, z_{ k m n}^{L, \rm in}\}$, which physically represent the  right-, left-handed circularly polarized components of the wave, respectively. The canonical symplectic structure $\Omega$ on this covariant phase space is
\bea \label{sympR}
\Omega(A_R^1,A_R^2)  :=\frac{1}{2} \int_\Sigma d\Sigma \, {\rm Re} \, (\overline{A^1_R}  H_R^2-A^2_R  \overline{H_R^1})\, ,
\eea
with $H_{R,a}:=n^b F^+_{ba}$, which verifies  $\Omega(A_R^1,A_R^2) = \Omega_0(A^1,A^2)$, thereby constant in time $t$. The  Noether charge associated to electric-magnetic duality rotations is
$
Q(A_R)=\int_{\Sigma} d\Sigma\,  {\rm Im} \,(A_{a,L} H^a_R) = \Omega(A_R,-i A_R) \, . \label{classicalQ} 
$
A straightforward calculation yields
\bea\label{classicalQ}
Q(A_R)= 2 \hbar \, R \sum_{n=0}^{\infty} \sum_{m=-\infty}^{\infty}\int_{-\infty}^{\infty} dk ( |z^{L, \rm in}_{kmn}|^2- |z^{R, \rm in}_{kmn}|^2)\, .
\eea
This is, for a given solution $A_R$ of the source-free Maxwell equations inside our waveguide, $Q(A_R)$ measures the net difference between right- and left-handed amplitudes over all $k$, $m$ and $n$. This is the familiar V Stokes parameter used in classical optics, which accounts for the circular polarization state of electromagnetic radiation.

The general solution given in (\ref{generalsolution}) for the electromagnetic potential inside the waveguide is tailored to the co-moving reference frame.  Equivalently, it can be recast in a form adapted to the inertial frame, which becomes particularly convenient at late times. In this reformulation, the mode expansion in  (\ref{generalsolution}) is reorganized as follows:
\bea \label{outrep}
A_a^R&=&\sum_{(hkmn)\in H^{+>}_h} z^{R,{\rm out}}_{hkmn} A^{R,{\rm out}}_{a, hkmn}+ \sum_{(hkmn)\in H^{->}_h} \overline{ z_{h k m n}^{L, {\rm out}}} \,\overline{  A^{L,{\rm out}}_{a, h k m n}}\, ,
\eea
where now  the circular polarization vectors are
\bea
 A_{a,hkmn}^{R,{\rm out}}\sim \frac{i R^2 }{j_{mn}^2}e^{-i\left[(h\omega_{kmn}-m\Omega_0-kv_0)\gamma t+(k-h\omega_{kmn} v+m v_0 \Omega_0)\gamma z+m\phi\right]}\times \nonumber\\
 \left[  \frac{j_{mn}}{R} J_{m+1}\left( \chi \right) {\bf \overline{ m_a}}{\bf e^{-i \Omega_0 \gamma(t-v_0z)}}-e^{-\theta} i (h\omega_{kmn}+k){\bf \Bell_a} J_{m}\left( \chi\right) \right]  \nonumber\, ,\\
 A_{a,hkmn}^{L,{\rm out}} \sim \frac{i R^2 }{j_{mn}^2}e^{-i\left[(h\omega_{kmn}-m\Omega_0-kv_0)\gamma t+(k-h\omega_{kmn} v+m v_0 \Omega_0)\gamma z+m\phi\right]}\times\nonumber\\
 \left[   \frac{j_{mn}}{R} J_{m-1}\left(  \chi\right)  {\bf m_a}{\bf e^{+i \Omega_0 \gamma (t-v_0z)}} + e^{-\theta} i (h\omega_{kmn}+k){\bf \Bell_a} J_{m}\left(  \chi\right) \right] \nonumber\, ,
\eea 
in the $t\to \infty$ limit, after switching to the inertial  coordinate system ($\chi\equiv\frac{j_{mn}}{R} \rho$). On the other hand 
\bea
H^{\pm \lessgtr}_h=\{(k,m,n)\in  \mathbb R\times \mathbb Z \times \mathbb Z \, /  \hspace{3cm}\\
   h \sqrt{k^2+j_{mn}^2/R^2}-m\Omega_0-kv_0 \pm \Omega_0 \lessgtr 0  \}\, , \nonumber
\eea
which satisfy $H^{\pm >}_h=H^{\mp <}_{-h}$. Notice how the rotating frame introduces the global factor $e^{-i\Omega_0 \tilde t}$ in the right-handed field modes. This is because the field  $A^R_a$ has a well-definite helicity, and rotates as a spin-1 field (while the left-handed field $A^L_a=\overline{A^R_a}$ has the opposite one).

\noindent\textit{\textbf{Quantization and anomaly: asymmetric photon pair creation}.}
In the quantum theory, the coupling with  the accelerating waveguide and the induced  time-dependent boundary conditions  are expected to excite  real photons  out of the quantum vacuum by the  dynamical casimir effect \cite{10.1063/1.1665432, Wilson:2011rsw}. If the final state of the waveguide has a preferred helicity direction, as in our case, one can further expect right- and left-handed photons to be created in different amounts. If this happens then $\langle Q\rangle$ cannot be constant in time anymore.

To obtain a quantum description of the electromagnetic field inside the waveguide we apply the standard algebraic quantization procedure.   Glossing over technical details, given a suitable algebra of observables satisfying the canonical commutation relations, a choice of a  vacuum state allows us to obtain a Hilbert  space representation  in terms of  operators,  via the GNS construction \cite{doi:10.1142/S0217751X13300238}. 
 As quantum field theories possess infinitely many degrees of freedom, many (possibly unitarily inequivalent) Hilbert space representations exist \cite{Wald:1995yp}. Mathematically, the problem of specifying a vacuum state can be reduced to that of selecting a suitable ``complex structure'' on the space of classical solutions  \cite{Ashtekar1975zn}, and in this case the resulting Hilbert space is a Fock space where vectors can be physically identified with photon states.

 A complex structure can be obtained  by decomposing the classical solutions $A_R$ in terms of positive- and negative-frequency modes with respect to some class of observers. We have previously identified two natural decompositions of this form  for the field. This is, {\it at early times} the decomposition  (\ref{generalsolution}) shows that the modes $A_{a,1kmn}^{R,\rm in}$ oscillate with positive values of frequency as measured by the time $t$ of inertial observers, while $\overline{A_{a,1kmn}^{L,\rm in}}$ have negative frequency. If we denote (\ref{generalsolution}) by $A_R=A_R^++A_R^-$, where $A_R^+$ is the positive part and $A_R^-$ the negative one, then we can define a linear map $J_{\rm in}: \Gamma_R\to \Gamma_R$ by $J_{\rm in}A_R^\pm=\pm i A_R^\pm$, so that $J_{\rm in}A_R=i A_R^+-i A_R^-$, and therefore  $J_{\rm in}^2=-\mathbb I$. This defines $J_{\rm in}$ as a complex structure, and the doublet $(\Gamma_R, J^{\rm in})$ becomes a complex vector space. It can be further shown that $J_{\rm in}$ is compatible with the symplectic structure, in the sense that $\Omega(J_{\rm in}A_R^1, J_{\rm in} A_R^2)=\Omega(A_R^1,A_R^2)$. This allows us to endow our phase space with an hermitian inner product $\langle A_R^1,A_R^2\rangle_{\rm in}=\frac{1}{2\hbar}\left[\Omega(A_R^1,J_{\rm in}A_R^2)+i\Omega(A_R^1,A_R^2) \right]$. The triplet $(\Gamma_R, J_{\rm in}, \langle \cdot , \cdot\rangle_{\rm in})$ gives us then the 1-photon Hilbert space of the theory (``1st quantization''), and from this one can directly obtain a bosonic Fock space $\mathbb F_{\rm in}$ describing multi-photon states (``2nd quantization''). The quantum electromagnetic potential  is  given by (\ref{generalsolution}) with the substitution $\{  z_{ k m n}^{R, \rm in}, \overline{ z_{ k m n}^{L, \rm in}}\}\to \{  a_{ k m n}^{R, \rm in},  a_{ k m n}^{L,\rm in, \dagger}\}$ in terms of right- (R) and left-handed (L) photon creation and annihilation operators, that satisfy the familiar canonical commutation relations. The ``in'' vacuum state is the ground state of our theory and satisfies $a^{R,\rm in}_{1 kmn}|\rm in\rangle = 0$.

On the other hand,  the decomposition  (\ref{outrep}) shows that, {\it at late times},  $A_{a,hkmn}^{R,\rm out}$ are  the modes that oscillate with positive frequency with respect to the time $t$ of the inertial reference frame, while $\overline{A_{a,hkmn}^{L,\rm out}}$ are of negative-frequency. By repeating exactly the same steps as before we end up with another Fock space $\mathbb F_{\rm out}$ at late times, with a different ``out'' vacuum state. The quantum electromagnetic potential is similarly  given by (\ref{outrep}) with the substitution $\{  z_{h k m n}^{R,\rm out}, \overline{ z_{h k m n}^{L,\rm out}}\}\to \{  a_{h k m n}^{R, \rm out},  a_{h k m n}^{L,\rm out, \dagger}\}$.

The expectation values of the operator $\hat Q$ representing  the Noether charge (\ref{classicalQ}) in the quantum theory produce ultraviolet divergences, a common feature that occurs for any observable that is quadratic in the fields.   To obtain physically sensible results we need to apply renormalization. Using standard prescriptions  for regularization (covariant point-splitting  \cite{Wald:1995yp}) one can get
\bea
\langle {\rm in}|\hat  Q (t\to+\infty) | {\rm in}\rangle-\langle {\rm in}|\hat  Q (t\to-\infty) | {\rm in}\rangle=\langle {\rm in}|  :\hat Q: | {\rm in}\rangle\, , \nonumber  
\eea
where $:\hat Q:$ denotes the result of evaluating (\ref{classicalQ}) after ``normal-ordering'' with respect to the ``out'' vacuum:
\bea
:\hat Q: \, = 2\hbar \, R \sum_{n=0}^{\infty} \sum_{m=-\infty}^{\infty}\int_{-\infty}^{\infty} dk (a^{L,\rm out\dagger}_{1kmn}a^{L, \rm out}_{1kmn}  - a^{R,\rm out\dagger}_{1kmn}a^{R, \rm out}_{1kmn})\, . \nonumber
\eea
Now, because the two Fock Hilbert spaces at early and late times are different, the ``in'' and ``out'' sets of creation/annihilation operators will differ. In particular, the in vacuum state is no longer the vacuum state of the out Fock space,  $a^{R,\rm out}_{1 kmn}|\rm in\rangle \neq 0$, but   is rather in correspondence with a multi-photon state. Consequently,  the expectation value  can be non-vanishing. After evaluating the Bogoliubov transformations between the two Fock spaces (to be published elsewhere), the VEV  reduces to
\bea
2\hbar \, R\left[   \sum_{\substack{\forall k'm'n' \\(hkmn)\in H^{+<}_{h}} }|\beta^R_{1k'm'n'\atop hkmn}|^2 - \sum_{\substack{\forall k'm'n' \\(hkmn)\in H^{-<}_{h} }}  |\beta^R_{-1k'm'n'\atop -hkmn}|^2 \right]\nonumber\, .
\eea
The first sum physically represents the amount of photon modes produced with helicity $+1$, while the second  denotes photons excited with helicity $-1$. Using  Bogoliubov identities one finds that this difference is non trivial:
\bea
\langle {\rm in}|  :\hat Q: | {\rm in}\rangle = -2\hbar \, R\, s \sum_{(kmn)\in H^{-s<}_1-H^{s<}_1 }1\, ,\hspace{0.5cm} \label{final}
\eea
with $s=  {\rm sign}\, \Omega_0$ and
\bea \label{set}
H^{-s<}_1-H^{s<}_1 =\Big\{(k,m,n)\in \mathbb R\times \mathbb Z\times \mathbb N\, /  \hspace{2cm}\\
   -|\Omega_0| < \sqrt{k^2+\frac{j_{mn}^2}{R^2}}-m|\Omega_0|-k v_0< |\Omega_0| \Big\} \, . \nonumber
\eea
This set is bounded, which makes (\ref{final}) finite. Further, it is not difficult to see that this set is non-empty provided that both $\Omega_0$ {\it and} $v_0$ are non-zero. Thus, we conclude that $\langle {\rm in}|\hat  Q (t) | {\rm in}\rangle$ {\it fails to be conserved in time}.

\noindent\textit{\textbf{Conclusions and future prospects}.} 
Our results reveal that the  electric-magnetic duality symmetry of the classical source-free Maxwell theory fails to survive  quantization inside an accelerating waveguide.  As  with the chiral fermion anomaly in 1+1 dimensions \cite{PhysRevD.108.105025}, a preferred helicity direction in the background  produces  an imbalance between right- and left-handed photons in the ``in'' vacuum state at late times. Ultimately, this asymmetry arises because more right- than left-handed electromagnetic modes (or viceversa) are allowed inside the waveguide at late times  as seen by inertial observers ($H^{+>}_h\neq H^{->}_h$ in \eqref{outrep}). Consequently, $\langle {\rm in}|\hat  Q (t\to+\infty) | {\rm in}\rangle\neq \langle {\rm in}|\hat  Q (t\to-\infty) | {\rm in}\rangle$, and the VEV is not constant in time. This is a genuinely quantum effect, with potentially observable consequences.  

Although  idealized, the model considered here serves as a proof-of-concept  and opens the door to studying the implications of this duality anomaly in the laboratory through analogue gravity models. However, experimental realization remains challenging.
 While the  precise mode counting in (\ref{final}) is technically difficult,  the  expected asymmetry between right- and left-handed photons  in this model can be  estimated of the order of $|\Delta N|\sim  |\frac{R \Omega_0 v_0}{\sqrt{1-v_0^2}}|$. In the ultra-relativistic limit, $|R\Omega_0|\sim 0.9$, $|v_0|\sim 0.9$, this yields $\Delta N\sim 2$ photons. Even in this optimal case, the number is small and may be hard to detect in the presence of ambient noise and other experimental limitations.

For models with other boundary conditions, we can have a time evolution of $Q_0(A)$ already at the classical level, since  electric-magnetic duality may not hold inside the waveguide. For example, with perfectly conducting plates, the standard ``Transverse Electric''  and ``Transverse Magnetic'' modes are not related by a duality transformation  because they possess different frequency spectra. This suggests that, in the quantum theory, equation (\ref{final}) will include two contributions: a classical one due to explicit symmetry breaking by the boundary conditions, and a genuinely quantum term from the anomaly.

To overcome  these limitations and make this quantum effect more accessible to experiment, we plan to investigate the potential enhancement achieved by working with squeezed  states. These states are widely used in quantum optics for their ability to amplify  quantum correlations  and mitigate suppression by environmental effects. Notably, they have recently been  shown to enhance stimulated particle creation and related quantum entanglement in analogue gravity setups \cite{PhysRevLett.128.091301, PhysRevD.106.105021}. Thus, they can have the potential to amplify the  expectation value of the operator $Q$.  We plan to explore  these ideas in future work.

{\bf \em Acknowledgments.} 
I thank Ivan Agullo,  Albert Ferrando and Jose Navarro-Salas for  useful discussions and feedback. 
I acknowledge support through {\it Atraccion de Talento Cesar Nombela} grant No 2023-T1/TEC-29023, funded by Comunidad de Madrid (Spain); as well as   financial support  via the Spanish Grant  PID2023-149560NB-C21, funded by MCIU/AEI/10.13039/501100011033/FEDER, UE.

\bibliography{references}

\begin{thebibliography}{28}%
\makeatletter
\providecommand \@ifxundefined [1]{%
 \@ifx{#1\undefined}
}%
\providecommand \@ifnum [1]{%
 \ifnum #1\expandafter \@firstoftwo
 \else \expandafter \@secondoftwo
 \fi
}%
\providecommand \@ifx [1]{%
 \ifx #1\expandafter \@firstoftwo
 \else \expandafter \@secondoftwo
 \fi
}%
\providecommand \natexlab [1]{#1}%
\providecommand \enquote  [1]{``#1''}%
\providecommand \bibnamefont  [1]{#1}%
\providecommand \bibfnamefont [1]{#1}%
\providecommand \citenamefont [1]{#1}%
\providecommand \href@noop [0]{\@secondoftwo}%
\providecommand \href [0]{\begingroup \@sanitize@url \@href}%
\providecommand \@href[1]{\@@startlink{#1}\@@href}%
\providecommand \@@href[1]{\endgroup#1\@@endlink}%
\providecommand \@sanitize@url [0]{\catcode `\\12\catcode `\$12\catcode
  `\&12\catcode `\#12\catcode `\^12\catcode `\_12\catcode `\%12\relax}%
\providecommand \@@startlink[1]{}%
\providecommand \@@endlink[0]{}%
\providecommand \url  [0]{\begingroup\@sanitize@url \@url }%
\providecommand \@url [1]{\endgroup\@href {#1}{\urlprefix }}%
\providecommand \urlprefix  [0]{URL }%
\providecommand \Eprint [0]{\href }%
\providecommand \doibase [0]{http://dx.doi.org/}%
\providecommand \selectlanguage [0]{\@gobble}%
\providecommand \bibinfo  [0]{\@secondoftwo}%
\providecommand \bibfield  [0]{\@secondoftwo}%
\providecommand \translation [1]{[#1]}%
\providecommand \BibitemOpen [0]{}%
\providecommand \bibitemStop [0]{}%
\providecommand \bibitemNoStop [0]{.\EOS\space}%
\providecommand \EOS [0]{\spacefactor3000\relax}%
\providecommand \BibitemShut  [1]{\csname bibitem#1\endcsname}%
\let\auto@bib@innerbib\@empty
\bibitem [{\citenamefont {Deser}\ and\ \citenamefont
  {Teitelboim}(1976)}]{PhysRevD.13.1592}%
  \BibitemOpen
  \bibfield  {author} {\bibinfo {author} {\bibfnamefont {S.}~\bibnamefont
  {Deser}}\ and\ \bibinfo {author} {\bibfnamefont {C.}~\bibnamefont
  {Teitelboim}},\ }\href {\doibase 10.1103/PhysRevD.13.1592} {\bibfield
  {journal} {\bibinfo  {journal} {Phys. Rev. D}\ }\textbf {\bibinfo {volume}
  {13}},\ \bibinfo {pages} {1592} (\bibinfo {year} {1976})}\BibitemShut
  {NoStop}%
\bibitem [{\citenamefont {Calkin}(1965)}]{10.1119/1.1971089}%
  \BibitemOpen
  \bibfield  {author} {\bibinfo {author} {\bibfnamefont {M.~G.}\ \bibnamefont
  {Calkin}},\ }\href {\doibase 10.1119/1.1971089} {\bibfield  {journal}
  {\bibinfo  {journal} {Am. J. Phys.}\ }\textbf {\bibinfo {volume} {33}},\
  \bibinfo {pages} {958} (\bibinfo {year} {1965})}\BibitemShut {NoStop}%
\bibitem [{\citenamefont {Agullo}\ \emph
  {et~al.}(2017{\natexlab{a}})\citenamefont {Agullo}, \citenamefont {del Rio},\
  and\ \citenamefont {Navarro-Salas}}]{PhysRevLett.118.111301}%
  \BibitemOpen
  \bibfield  {author} {\bibinfo {author} {\bibfnamefont {I.}~\bibnamefont
  {Agullo}}, \bibinfo {author} {\bibfnamefont {A.}~\bibnamefont {del Rio}}, \
  and\ \bibinfo {author} {\bibfnamefont {J.}~\bibnamefont {Navarro-Salas}},\
  }\href {\doibase 10.1103/PhysRevLett.118.111301} {\bibfield  {journal}
  {\bibinfo  {journal} {Phys. Rev. Lett.}\ }\textbf {\bibinfo {volume} {118}},\
  \bibinfo {pages} {111301} (\bibinfo {year} {2017}{\natexlab{a}})}\BibitemShut
  {NoStop}%
\bibitem [{\citenamefont {Agullo}\ \emph
  {et~al.}(2018{\natexlab{a}})\citenamefont {Agullo}, \citenamefont {del Rio},\
  and\ \citenamefont {Navarro-Salas}}]{PhysRevD.98.125001}%
  \BibitemOpen
  \bibfield  {author} {\bibinfo {author} {\bibfnamefont {I.}~\bibnamefont
  {Agullo}}, \bibinfo {author} {\bibfnamefont {A.}~\bibnamefont {del Rio}}, \
  and\ \bibinfo {author} {\bibfnamefont {J.}~\bibnamefont {Navarro-Salas}},\
  }\href {\doibase 10.1103/PhysRevD.98.125001} {\bibfield  {journal} {\bibinfo
  {journal} {Phys. Rev. D}\ }\textbf {\bibinfo {volume} {98}},\ \bibinfo
  {pages} {125001} (\bibinfo {year} {2018}{\natexlab{a}})}\BibitemShut
  {NoStop}%
\bibitem [{\citenamefont {Adler}(1969)}]{PhysRev.177.2426}%
  \BibitemOpen
  \bibfield  {author} {\bibinfo {author} {\bibfnamefont {S.~L.}\ \bibnamefont
  {Adler}},\ }\href {\doibase 10.1103/PhysRev.177.2426} {\bibfield  {journal}
  {\bibinfo  {journal} {Phys. Rev.}\ }\textbf {\bibinfo {volume} {177}},\
  \bibinfo {pages} {2426} (\bibinfo {year} {1969})}\BibitemShut {NoStop}%
\bibitem [{\citenamefont {{Bell}}\ and\ \citenamefont
  {{Jackiw}}(1969)}]{1969NCimA..60...47B}%
  \BibitemOpen
  \bibfield  {author} {\bibinfo {author} {\bibfnamefont {J.~S.}\ \bibnamefont
  {{Bell}}}\ and\ \bibinfo {author} {\bibfnamefont {R.}~\bibnamefont
  {{Jackiw}}},\ }\href {\doibase 10.1007/BF02823296} {\bibfield  {journal}
  {\bibinfo  {journal} {Nuovo Cimento A Serie}\ }\textbf {\bibinfo {volume}
  {60}},\ \bibinfo {pages} {47} (\bibinfo {year} {1969})}\BibitemShut {NoStop}%
\bibitem [{\citenamefont {del Rio}\ \emph {et~al.}(2020)\citenamefont {del
  Rio}, \citenamefont {Sanchis-Gual}, \citenamefont {Mewes}, \citenamefont
  {Agullo}, \citenamefont {Font},\ and\ \citenamefont
  {Navarro-Salas}}]{PhysRevLett.124.211301}%
  \BibitemOpen
  \bibfield  {author} {\bibinfo {author} {\bibfnamefont {A.}~\bibnamefont {del
  Rio}}, \bibinfo {author} {\bibfnamefont {N.}~\bibnamefont {Sanchis-Gual}},
  \bibinfo {author} {\bibfnamefont {V.}~\bibnamefont {Mewes}}, \bibinfo
  {author} {\bibfnamefont {I.}~\bibnamefont {Agullo}}, \bibinfo {author}
  {\bibfnamefont {J.~A.}\ \bibnamefont {Font}}, \ and\ \bibinfo {author}
  {\bibfnamefont {J.}~\bibnamefont {Navarro-Salas}},\ }\href {\doibase
  10.1103/PhysRevLett.124.211301} {\bibfield  {journal} {\bibinfo  {journal}
  {Phys. Rev. Lett.}\ }\textbf {\bibinfo {volume} {124}},\ \bibinfo {pages}
  {211301} (\bibinfo {year} {2020})}\BibitemShut {NoStop}%
\bibitem [{\citenamefont {del Rio}(2021)}]{PhysRevD.104.065012}%
  \BibitemOpen
  \bibfield  {author} {\bibinfo {author} {\bibfnamefont {A.}~\bibnamefont {del
  Rio}},\ }\href {\doibase 10.1103/PhysRevD.104.065012} {\bibfield  {journal}
  {\bibinfo  {journal} {Phys. Rev. D}\ }\textbf {\bibinfo {volume} {104}},\
  \bibinfo {pages} {065012} (\bibinfo {year} {2021})}\BibitemShut {NoStop}%
\bibitem [{\citenamefont {Sanchis-Gual}\ and\ \citenamefont {del
  Rio}(2023)}]{PhysRevD.108.044052}%
  \BibitemOpen
  \bibfield  {author} {\bibinfo {author} {\bibfnamefont {N.}~\bibnamefont
  {Sanchis-Gual}}\ and\ \bibinfo {author} {\bibfnamefont {A.}~\bibnamefont {del
  Rio}},\ }\href {\doibase 10.1103/PhysRevD.108.044052} {\bibfield  {journal}
  {\bibinfo  {journal} {Phys. Rev. D}\ }\textbf {\bibinfo {volume} {108}},\
  \bibinfo {pages} {044052} (\bibinfo {year} {2023})}\BibitemShut {NoStop}%
\bibitem [{\citenamefont {Leong}\ \emph {et~al.}(2025)\citenamefont {Leong},
  \citenamefont {Tome}, \citenamefont {Bustillo}, \citenamefont {del Rio},\
  and\ \citenamefont
  {Sanchis-Gual}}]{leong2025gravitationalwavesignaturesmirrorasymmetry}%
  \BibitemOpen
  \bibfield  {author} {\bibinfo {author} {\bibfnamefont {S.~H.~W.}\
  \bibnamefont {Leong}}, \bibinfo {author} {\bibfnamefont {A.~F.}\ \bibnamefont
  {Tome}}, \bibinfo {author} {\bibfnamefont {J.~C.}\ \bibnamefont {Bustillo}},
  \bibinfo {author} {\bibfnamefont {A.}~\bibnamefont {del Rio}}, \ and\
  \bibinfo {author} {\bibfnamefont {N.}~\bibnamefont {Sanchis-Gual}},\ }\href
  {https://arxiv.org/abs/2501.11663} {} (\bibinfo {year} {2025}),\ \Eprint
  {http://arxiv.org/abs/2501.11663} {arXiv:2501.11663 [gr-qc]} \BibitemShut
  {NoStop}%
\bibitem [{\citenamefont {Agullo}\ \emph
  {et~al.}(2017{\natexlab{b}})\citenamefont {Agullo}, \citenamefont {del Rio},\
  and\ \citenamefont {Navarro-Salas}}]{doi:10.1142/S0218271817420019}%
  \BibitemOpen
  \bibfield  {author} {\bibinfo {author} {\bibfnamefont {I.}~\bibnamefont
  {Agullo}}, \bibinfo {author} {\bibfnamefont {A.}~\bibnamefont {del Rio}}, \
  and\ \bibinfo {author} {\bibfnamefont {J.}~\bibnamefont {Navarro-Salas}},\
  }\href {\doibase 10.1142/S0218271817420019} {\bibfield  {journal} {\bibinfo
  {journal} {Int. J. Mod. Phys. D}\ }\textbf {\bibinfo {volume} {26}},\
  \bibinfo {pages} {1742001} (\bibinfo {year}
  {2017}{\natexlab{b}})}\BibitemShut {NoStop}%
\bibitem [{\citenamefont {Agullo}\ \emph
  {et~al.}(2018{\natexlab{b}})\citenamefont {Agullo}, \citenamefont {del Rio},\
  and\ \citenamefont {Navarro-Salas}}]{sym10120763}%
  \BibitemOpen
  \bibfield  {author} {\bibinfo {author} {\bibfnamefont {I.}~\bibnamefont
  {Agullo}}, \bibinfo {author} {\bibfnamefont {A.}~\bibnamefont {del Rio}}, \
  and\ \bibinfo {author} {\bibfnamefont {J.}~\bibnamefont {Navarro-Salas}},\
  }\href {\doibase 10.3390/sym10120763} {\bibfield  {journal} {\bibinfo
  {journal} {Symmetry}\ }\textbf {\bibinfo {volume} {10}} (\bibinfo {year}
  {2018}{\natexlab{b}}),\ 10.3390/sym10120763}\BibitemShut {NoStop}%
\bibitem [{\citenamefont {Dirac}(1931)}]{Dirac1931kp}%
  \BibitemOpen
  \bibfield  {author} {\bibinfo {author} {\bibfnamefont {P.~A.~M.}\
  \bibnamefont {Dirac}},\ }\href {\doibase 10.1098/rspa.1931.0130} {\bibfield
  {journal} {\bibinfo  {journal} {Proc. Roy. Soc. Lond. A}\ }\textbf {\bibinfo
  {volume} {133}},\ \bibinfo {pages} {60} (\bibinfo {year} {1931})}\BibitemShut
  {NoStop}%
\bibitem [{\citenamefont {Seiberg}\ and\ \citenamefont
  {Witten}(1994)}]{SEIBERG199419}%
  \BibitemOpen
  \bibfield  {author} {\bibinfo {author} {\bibfnamefont {N.}~\bibnamefont
  {Seiberg}}\ and\ \bibinfo {author} {\bibfnamefont {E.}~\bibnamefont
  {Witten}},\ }\href {\doibase https://doi.org/10.1016/0550-3213(94)90124-4}
  {\bibfield  {journal} {\bibinfo  {journal} {Nucl. Phys. B}\ }\textbf
  {\bibinfo {volume} {426}},\ \bibinfo {pages} {19} (\bibinfo {year}
  {1994})}\BibitemShut {NoStop}%
\bibitem [{\citenamefont {Hawking}(1975)}]{cmp/1103899181}%
  \BibitemOpen
  \bibfield  {author} {\bibinfo {author} {\bibfnamefont {S.~W.}\ \bibnamefont
  {Hawking}},\ }\href {https://link.springer.com/article/10.1007/BF02345020}
  {\bibfield  {journal} {\bibinfo  {journal} {Commun. Math. Phys.}\ }\textbf
  {\bibinfo {volume} {43}},\ \bibinfo {pages} {199 } (\bibinfo {year}
  {1975})}\BibitemShut {NoStop}%
\bibitem [{\citenamefont {Barcelo}\ \emph {et~al.}(2005)\citenamefont
  {Barcelo}, \citenamefont {Liberati},\ and\ \citenamefont
  {Visser}}]{Barcel__2005}%
  \BibitemOpen
  \bibfield  {author} {\bibinfo {author} {\bibfnamefont {C.}~\bibnamefont
  {Barcelo}}, \bibinfo {author} {\bibfnamefont {S.}~\bibnamefont {Liberati}}, \
  and\ \bibinfo {author} {\bibfnamefont {M.}~\bibnamefont {Visser}},\
  }\href@noop {} {\bibfield  {journal} {\bibinfo  {journal} {Living Reviews in
  Relativity}\ }\textbf {\bibinfo {volume} {8}} (\bibinfo {year}
  {2005})}\BibitemShut {NoStop}%
\bibitem [{\citenamefont {Lee}\ and\ \citenamefont
  {Wald}(1990)}]{10.1063/1.528801}%
  \BibitemOpen
  \bibfield  {author} {\bibinfo {author} {\bibfnamefont {J.}~\bibnamefont
  {Lee}}\ and\ \bibinfo {author} {\bibfnamefont {R.~M.}\ \bibnamefont {Wald}},\
  }\href {\doibase 10.1063/1.528801} {\bibfield  {journal} {\bibinfo  {journal}
  {J. Math. Phys.}\ }\textbf {\bibinfo {volume} {31}},\ \bibinfo {pages} {725}
  (\bibinfo {year} {1990})}\BibitemShut {NoStop}%
\bibitem [{\citenamefont {Margalef-Bentabol}\ and\ \citenamefont
  {Villase\~nor}(2021)}]{PhysRevD.103.025011}%
  \BibitemOpen
  \bibfield  {author} {\bibinfo {author} {\bibfnamefont {J.}~\bibnamefont
  {Margalef-Bentabol}}\ and\ \bibinfo {author} {\bibfnamefont {E.~J.~S.}\
  \bibnamefont {Villase\~nor}},\ }\href {\doibase 10.1103/PhysRevD.103.025011}
  {\bibfield  {journal} {\bibinfo  {journal} {Phys. Rev. D}\ }\textbf {\bibinfo
  {volume} {103}},\ \bibinfo {pages} {025011} (\bibinfo {year}
  {2021})}\BibitemShut {NoStop}%
\bibitem [{\citenamefont {Torres~del Castillo}(2003)}]{Torres2003}%
  \BibitemOpen
  \bibfield  {author} {\bibinfo {author} {\bibfnamefont {G.~M.}\ \bibnamefont
  {Torres~del Castillo}},\ }\href {\doibase
  10.7208/chicago/9780226870373.001.0001} {\emph {\bibinfo {title} {{3D
  spinors, Spin-Weighted Functions and their Applications}}}}\ (\bibinfo
  {publisher} {Birkhauser},\ \bibinfo {address} {Boston, USA},\ \bibinfo {year}
  {2003})\BibitemShut {NoStop}%
\bibitem [{\citenamefont {Ashtekar}(1986)}]{10.1063/1.527187}%
  \BibitemOpen
  \bibfield  {author} {\bibinfo {author} {\bibfnamefont {A.}~\bibnamefont
  {Ashtekar}},\ }\href {\doibase 10.1063/1.527187} {\bibfield  {journal}
  {\bibinfo  {journal} {J. Math. Phys.}\ }\textbf {\bibinfo {volume} {27}},\
  \bibinfo {pages} {824} (\bibinfo {year} {1986})}\BibitemShut {NoStop}%
\bibitem [{\citenamefont {Moore}(1970)}]{10.1063/1.1665432}%
  \BibitemOpen
  \bibfield  {author} {\bibinfo {author} {\bibfnamefont {G.~T.}\ \bibnamefont
  {Moore}},\ }\href {\doibase 10.1063/1.1665432} {\bibfield  {journal}
  {\bibinfo  {journal} {J. Math. Phys.}\ }\textbf {\bibinfo {volume} {11}},\
  \bibinfo {pages} {2679} (\bibinfo {year} {1970})}\BibitemShut {NoStop}%
\bibitem [{\citenamefont {Wilson}\ \emph {et~al.}(2011)\citenamefont {Wilson},
  \citenamefont {Johansson}, \citenamefont {Pourkabirian}, \citenamefont
  {Simoen}, \citenamefont {Johansson}, \citenamefont {Duty}, \citenamefont
  {Nori},\ and\ \citenamefont {Delsing}}]{Wilson:2011rsw}%
  \BibitemOpen
  \bibfield  {author} {\bibinfo {author} {\bibfnamefont {C.~M.}\ \bibnamefont
  {Wilson}}, \bibinfo {author} {\bibfnamefont {G.}~\bibnamefont {Johansson}},
  \bibinfo {author} {\bibfnamefont {A.}~\bibnamefont {Pourkabirian}}, \bibinfo
  {author} {\bibfnamefont {M.}~\bibnamefont {Simoen}}, \bibinfo {author}
  {\bibfnamefont {J.~R.}\ \bibnamefont {Johansson}}, \bibinfo {author}
  {\bibfnamefont {T.}~\bibnamefont {Duty}}, \bibinfo {author} {\bibfnamefont
  {F.}~\bibnamefont {Nori}}, \ and\ \bibinfo {author} {\bibfnamefont
  {P.}~\bibnamefont {Delsing}},\ }\href {\doibase 10.1038/nature10561}
  {\bibfield  {journal} {\bibinfo  {journal} {Nature}\ }\textbf {\bibinfo
  {volume} {479}},\ \bibinfo {pages} {376} (\bibinfo {year}
  {2011})}\BibitemShut {NoStop}%
\bibitem [{\citenamefont {Benini}\ \emph {et~al.}(2013)\citenamefont {Benini},
  \citenamefont {Dappiaggi},\ and\ \citenamefont
  {Hack}}]{doi:10.1142/S0217751X13300238}%
  \BibitemOpen
  \bibfield  {author} {\bibinfo {author} {\bibfnamefont {M.}~\bibnamefont
  {Benini}}, \bibinfo {author} {\bibfnamefont {C.}~\bibnamefont {Dappiaggi}}, \
  and\ \bibinfo {author} {\bibfnamefont {T.-P.}\ \bibnamefont {Hack}},\ }\href
  {\doibase 10.1142/S0217751X13300238} {\bibfield  {journal} {\bibinfo
  {journal} {Int. J. Mod. Phys. A}\ }\textbf {\bibinfo {volume} {28}},\
  \bibinfo {pages} {1330023} (\bibinfo {year} {2013})}\BibitemShut {NoStop}%
\bibitem [{\citenamefont {Wald}(1995)}]{Wald:1995yp}%
  \BibitemOpen
  \bibfield  {author} {\bibinfo {author} {\bibfnamefont {R.~M.}\ \bibnamefont
  {Wald}},\ }\href
  {https://press.uchicago.edu/ucp/books/book/chicago/Q/bo3684008.html} {\emph
  {\bibinfo {title} {{Quantum Field Theory in Curved Space-Time and Black Hole
  Thermodynamics}}}},\ Chicago Lectures in Physics\ (\bibinfo  {publisher}
  {University of Chicago Press},\ \bibinfo {address} {Chicago, IL},\ \bibinfo
  {year} {1995})\BibitemShut {NoStop}%
\bibitem [{\citenamefont {Ashtekar}\ and\ \citenamefont
  {Magnon}(1975)}]{Ashtekar1975zn}%
  \BibitemOpen
  \bibfield  {author} {\bibinfo {author} {\bibfnamefont {A.}~\bibnamefont
  {Ashtekar}}\ and\ \bibinfo {author} {\bibfnamefont {A.}~\bibnamefont
  {Magnon}},\ }\href {\doibase 10.1098/rspa.1975.0181} {\bibfield  {journal}
  {\bibinfo  {journal} {Proc. Roy. Soc. Lond. A}\ }\textbf {\bibinfo {volume}
  {346}},\ \bibinfo {pages} {375} (\bibinfo {year} {1975})}\BibitemShut
  {NoStop}%
\bibitem [{\citenamefont {del R\'{\i}o}\ and\ \citenamefont
  {Agullo}(2023)}]{PhysRevD.108.105025}%
  \BibitemOpen
  \bibfield  {author} {\bibinfo {author} {\bibfnamefont {A.}~\bibnamefont {del
  R\'{\i}o}}\ and\ \bibinfo {author} {\bibfnamefont {I.}~\bibnamefont
  {Agullo}},\ }\href {\doibase 10.1103/PhysRevD.108.105025} {\bibfield
  {journal} {\bibinfo  {journal} {Phys. Rev. D}\ }\textbf {\bibinfo {volume}
  {108}},\ \bibinfo {pages} {105025} (\bibinfo {year} {2023})}\BibitemShut
  {NoStop}%
\bibitem [{\citenamefont {Agullo}\ \emph {et~al.}(2022)\citenamefont {Agullo},
  \citenamefont {Brady},\ and\ \citenamefont
  {Kranas}}]{PhysRevLett.128.091301}%
  \BibitemOpen
  \bibfield  {author} {\bibinfo {author} {\bibfnamefont {I.}~\bibnamefont
  {Agullo}}, \bibinfo {author} {\bibfnamefont {A.~J.}\ \bibnamefont {Brady}}, \
  and\ \bibinfo {author} {\bibfnamefont {D.}~\bibnamefont {Kranas}},\ }\href
  {\doibase 10.1103/PhysRevLett.128.091301} {\bibfield  {journal} {\bibinfo
  {journal} {Phys. Rev. Lett.}\ }\textbf {\bibinfo {volume} {128}},\ \bibinfo
  {pages} {091301} (\bibinfo {year} {2022})}\BibitemShut {NoStop}%
\bibitem [{\citenamefont {Brady}\ \emph {et~al.}(2022)\citenamefont {Brady},
  \citenamefont {Agullo},\ and\ \citenamefont {Kranas}}]{PhysRevD.106.105021}%
  \BibitemOpen
  \bibfield  {author} {\bibinfo {author} {\bibfnamefont {A.~J.}\ \bibnamefont
  {Brady}}, \bibinfo {author} {\bibfnamefont {I.}~\bibnamefont {Agullo}}, \
  and\ \bibinfo {author} {\bibfnamefont {D.}~\bibnamefont {Kranas}},\ }\href
  {\doibase 10.1103/PhysRevD.106.105021} {\bibfield  {journal} {\bibinfo
  {journal} {Phys. Rev. D}\ }\textbf {\bibinfo {volume} {106}},\ \bibinfo
  {pages} {105021} (\bibinfo {year} {2022})}\BibitemShut {NoStop}%
\end{thebibliography}%

\end{document}